%% file: main.tex
\documentclass[sigconf]{acmart}
\input{macros.tex}

\AtBeginDocument{%
  }

\copyrightyear{2024}
\acmYear{2024}
\setcopyright{rightsretained}
\acmConference[CIKM '24]{Proceedings of the 33rd ACM International Conference on Information and Knowledge Management}{October 21--25, 2024}{Boise, ID, USA}
\acmBooktitle{Proceedings of the 33rd ACM International Conference on Information and Knowledge Management (CIKM '24), October 21--25, 2024, Boise, ID, USA}
\acmDOI{10.1145/3627673.3679786}
\acmISBN{979-8-4007-0436-9/24/10}

\makeatletter
\gdef\@copyrightpermission{
   \begin{minipage}{0.3\columnwidth}
     \href{https://creativecommons.org/licenses/by-nc-sa/4.0/}{\includegraphics[width=0.90\textwidth]{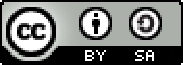}}
   \end{minipage}\hfill
   \begin{minipage}{0.7\columnwidth}
     \href{https://creativecommons.org/licenses/by-nc-sa/4.0/}{This work is licensed under a Creative Commons Attribution-ShareAlike International 4.0 License.}
   \end{minipage}
   \vspace{5pt}
}
\makeatother

\begin{document}

\title[Link Polarity Prediction from Sparse and Noisy Labels]{Link Polarity Prediction from Sparse and Noisy Labels via Multiscale Social Balance}

\author{Marco Minici}
\authornote{Contact author} 
\email{marco.minici@icar.cnr.it}
\orcid{0000-0002-9641-8916}
\affiliation{%
  \institution{University of Pisa, Pisa, Italy}
  \city{}
  \country{}
  \institution{ICAR-CNR, Rende, Italy}
  \city{}
  \country{}
}

\author{Federico Cinus}
\email{federico.cinus@centai.eu}
\orcid{0000-0002-6696-9637}
\affiliation{%
  \institution{Sapienza University, Rome, Italy}
  \city{}
  \country{}
  \institution{CENTAI, Turin, Italy}
  \city{}
  \country{}
}

\author{Francesco Bonchi}
\email{bonchi@centai.eu}
\orcid{1234-5678-9012}
\affiliation{%
  \institution{CENTAI, Turin, Italy}
  \city{}
  \country{}
  \institution{Eurecat, Barcelona, Spain}
  \city{}
  \country{}
}

\author{Giuseppe Manco}
\email{giuseppe.manco@icar.cnr.it}
\orcid{0000-0001-9672-3833}
\affiliation{%
  \institution{ICAR-CNR, Rende, Italy}
  \city{}
  \country{}
}
\renewcommand{\shortauthors}{Marco Minici, Federico Cinus, Francesco Bonchi, \& Giuseppe Manco}

\begin{abstract}
 Signed Graph Neural Networks (SGNNs) have recently gained attention as an effective tool for several learning tasks on signed networks, i.e., graphs where edges have an associated polarity. One of these tasks is to predict the polarity of the links for which this information is missing, starting from the network structure and the other available polarities. However, when the available polarities are few and potentially noisy, such a task becomes challenging.

In this work, we devise a semi-supervised learning framework that builds around the novel concept of \emph{multiscale social balance} to improve the prediction of link polarities in settings characterized by limited data quantity and quality.
Our model-agnostic approach can seamlessly integrate with any SGNN architecture, dynamically reweighting the importance of each data sample while making strategic use of the structural information from unlabeled edges combined with social balance theory.

Empirical validation demonstrates that our approach outperforms established baseline models, effectively addressing the limitations imposed by noisy and sparse data. This result underlines the benefits of incorporating multiscale social balance into SGNNs, opening new avenues for robust and accurate predictions in signed network analysis.
\end{abstract}

\maketitle
\section{Introduction}
\label{sec:intro}
\input{intro.tex}

\section{Related Work}
\label{sec:related}
\input{related.tex}

\section{Problem Definition}
\label{sec:problem}
\input{problem.tex}

\section{Proposed Framework}
\label{sec:method}
\input{method.tex}

\section{Experiments}
\label{sec:experiments}
\input{experiments.tex}

\section{Conclusions}
\label{sec:conclusions}
\input{conclusions}

\vspace{-0.35cm}
\begin{acks}
MM and GM acknowledge partial support
by $(i)$ the SERICS project (PE00000014) under the NRRP MUR program funded by the EU - NGEU and $(ii)$ European Union - NextGenerationEU - National Recovery and Resilience Plan Project: “SoBig-Data.it - Strengthening the Italian RI for Social Mining and Big Data Analytics” - Prot. IR0000013.
\end{acks}

\bibliographystyle{ACM-Reference-Format}
\balance
\bibliography{references}

\end{document}

%% file: macros.tex
\usepackage{amsmath}
\usepackage{url}
\usepackage{booktabs}
\usepackage{bm}
\usepackage{float}
\usepackage[ruled,linesnumbered]{algorithm2e} %
\usepackage[acronym]{glossaries}
\usepackage[show]{chato-notes}
\usepackage{multirow}
\usepackage{framed}
\usepackage{xcolor}
\usepackage{balance}

\graphicspath{{./figs/}}
\newacronym{gnn}{GNN}{Graph Neural Network}
\newacronym{sb}{SB}{Social Balance}
\newacronym{sbt}{SBT}{Social Balance Theory}
\newacronym{sgnn}{SGNN}{Signed Graph Neural Network}

\newenvironment{problemframe}{%
  \MakeFramed{\advance\hsize-\width \FrameRestore}%
  \vskip 1mm
}{%
  \vskip 1mm
  \endMakeFramed
}

\newtheorem{problem}{Problem}

\newcommand{\spara}[1]{\smallskip\noindent{\bf #1}}

\def\repo{https://github.com/mminici/SGNNfromSparseAndNoisyLabels}
\def\ourProcedure{\texttt{Learn2ReWeightSB}}

%% file: intro.tex
Signed networks have emerged as a simple yet powerful representation of social interactions:
vertices represent entities and edges between vertices represent interactions among them,
which can be friendly (positive) or antagonistic (negative)~\cite{harary1953notion}.
Many methods have thus been proposed to mine and analyse signed networks \cite{leskovec2010signed,TangCAL16,girdhar2017signed} in order to tackle various tasks such as clustering and community detection~\cite{zhao2020community, sun2020stable, tzeng2020discovering}, node classification~\cite{tang2016node}, signed link prediction and link sign prediction~\cite{yuan2017sne, chen2018bridge, javari2020rose, huang2022pole}.
The applications are many and diverse, ranging
from modeling interactions, polarization, and political stance in social media~\cite{KunegisLB09,bonchi2019discovering}, %
 mining user reviews~\cite{beigi2016signed},
 studying information diffusion and epidemics~\cite{li2013influence},
 recommending products in e-commerce sites~\cite{ma2009learning,victor2011trust}, and  estimating the structural balance of a (physical) complex system~\cite{antal2006,marvel2009}.

The success of Graph Neural Networks (GNNs)~\cite{kipf2016semi,velivckovic2018graph} has sparked interest in adapting the same message-passing paradigm to the domain of signed graphs. Consequently, many \emph{Signed Graph Neural Networks} (SGNNs) proposals have emerged~\cite{derr2018signed,huang2019signed,liu2021signed,huang2021sdgnn,singh2022signed,li2023signed}, making SGNN the dominant paradigm especially for what concerns link polarity\footnote{All over the paper, we use link polarity and link sign interchangeably.} prediction, a task in which they outperform other signed network embedding methods.
However, training SGNNs for state-of-the-art node representation learning often requires large amounts of high-quality data. As the quantity of data increases, maintaining its quality becomes increasingly challenging, since collecting the signs of all the links can be costly or impractical. This trade-off between data quantity and quality poses significant obstacles, making it difficult to achieve optimal model performance without incurring substantial data collection costs.
Moreover, as observed in recent studies \cite{zhang2023rsgnn,efstratiou2023non}, the collection of the links signs might be affected by noise for various reasons. There is thus a need for robust signed graph representation learning \cite{zhang2024signed}, to effectively train SGNNs in spite of sparse and noisy polarity information.

In this work, we tackle this problem by proposing a semi-supervised learning framework that builds around the concept of \emph{multiscale social balance} to improve the prediction of link polarities in settings characterized by limited data quantity and quality.

\spara{Multiscale social balance.}
Recently, many SGNNs have been proposed that leverage the classic social balance (SB) theory~\cite{derr2018signed,huang2019signed,huang2021sdgnn,Shu21,kim2023trustsgcn}.
The concept of social balance originates from the field of psychology \cite{heider1946attitudes} and states that higher-order relationships tend to form \emph{balanced triads} in which the product of the signs tends to be positive, reflecting a harmonious or balanced state: \emph{``a friend of my friend is my friend''}, \emph{``a foe of my foe is my friend''}, and so on.
While traditionally such a notion has been used to model undirected relationships,
recently social balance theory has been extended to directed graphs, allowing for a more nuanced understanding of interpersonal and group dynamics within these networks~\cite{dinh2020structural,aref2020multilevel,hao2024proper}.

Real-world data reveal that SB is predominant but not universally observed in large-scale networks \cite{leskovec2010signed}, highlighting the importance of discerning when and how to trust predictions based on this theory.
This recognition has led to advancements in the methodological assessment of social balance, such as the development of null models that preserve both network topology and signed degree sequences \cite{hao2024proper}. Furthermore, recent advancements underscore the necessity for a \emph{multiscale approach} to SB~\cite{aref2020multilevel}, capturing the complexity of social dynamics at multiple levels.

\begin{figure}[t!]
  \centering
  \begin{tabular}{cc}
    \includegraphics[width=.4\linewidth]{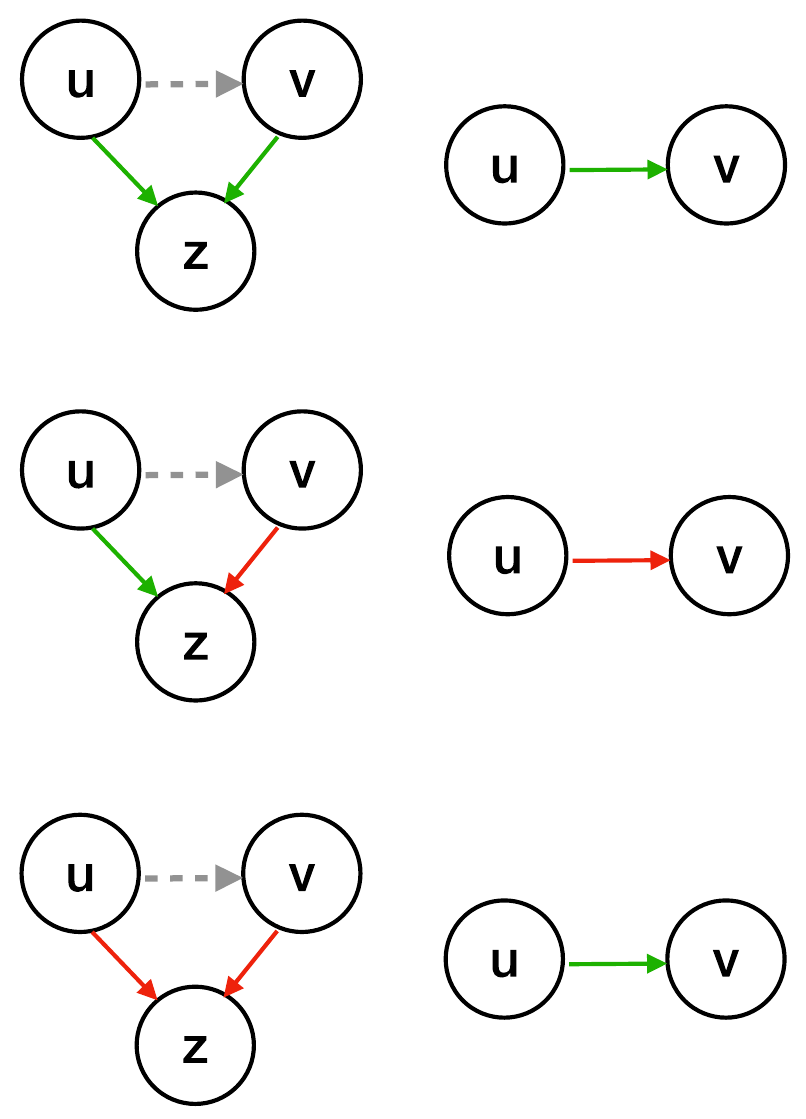}
    \label{fig:micro-scale-sb-explain}
    &
    \includegraphics[width=.4\linewidth]{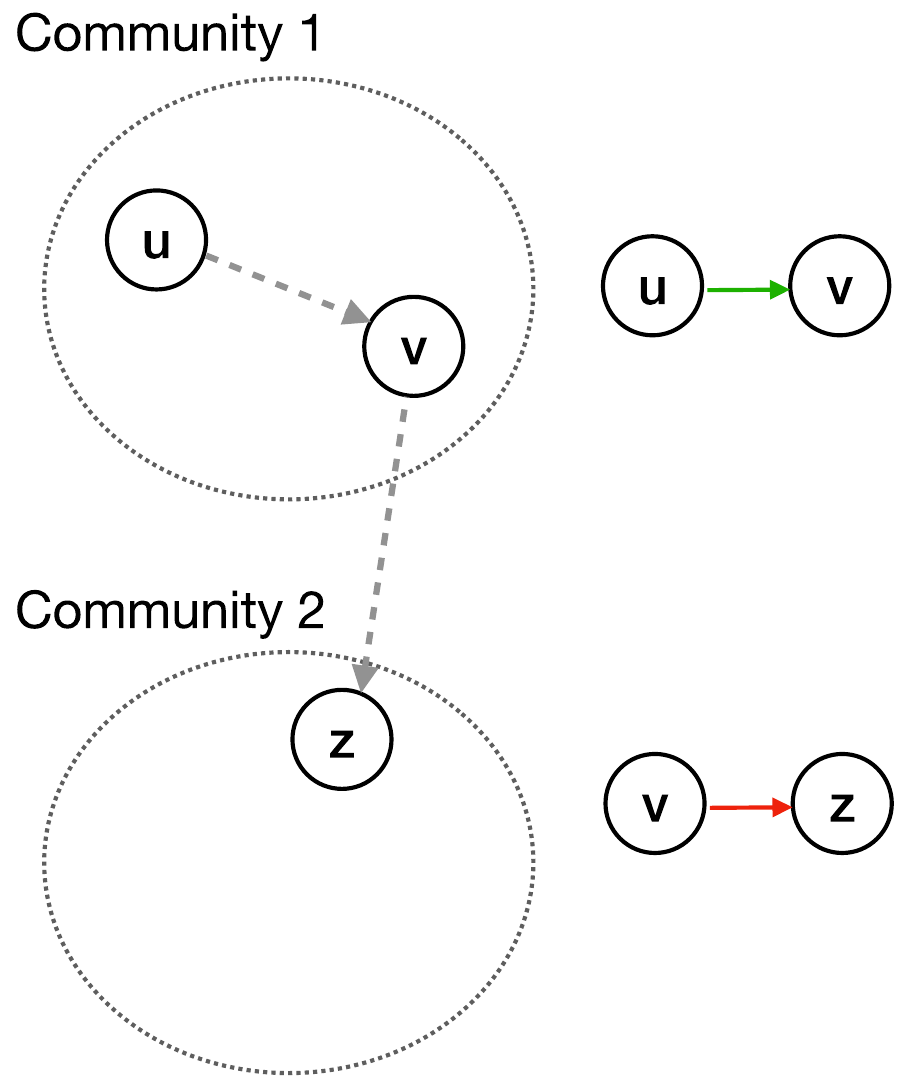}
    \label{fig:meso-scale-sb-explain}\\
    \small (a) Microscale Social Balance & \small (b) Mesoscale Social Balance
  \end{tabular}
  \vspace{-2mm}
  \caption{Depiction illustrating the use of (a) Microscale and (b) Mesoscale Social Balance to inform the link polarity prediction task. When closing a triad (a), the product of edge signs must always be equal to 1 (pos=1, neg=-1). When an edge crosses community boundaries its sign must be negative, while we assume it is positive inside the community (b).}
  \label{fig:social-balance-explain}
  \vspace{-3mm}
\end{figure}

Fig.~\ref{fig:social-balance-explain} depicts the key idea of
multiscale social balance (MSB) for predicting links sign. At the microscale, we exploit the classic social balance theory to predict a missing sign in a triad, so that the product of the signs is positive. MSB, however, also considers the mesoscale level, i.e., the existence of communities, and observes that communities are often antagonistic towards each other. Thus, when an edge crosses community boundaries its sign must be negative, while we assume it is positive inside the community.

\spara{Our contributions.} Following the intuition illustrated above, in this paper we present a semi-supervised framework to enhance the accuracy and robustness of link polarity predictions in SGNNs, in settings characterized by limited data quantity and quality that exploits multiscale social balance. The proposed framework dynamically reweights the importance of each data sample, effectively utilizing both labeled and unlabeled data to improve learning outcomes. Our approach is model-agnostic, as it can seamlessly integrate with any SGNN architecture.

We validate our proposal through extensive experimentation on four social signed graphs. We first train an SGNN model (SDGNN~\cite{huang2021sdgnn}) with and without our framework, observing improvements of up to 17.6\% in Accuracy and 14.2\% in Macro-F1 in a sparse and noisy data regime.
Next, we compare our framework against three semi-supervised baselines based on Pseudo-Labeling and a robust link polarity prediction method~\cite{zhang2023rsgnn}.
Our approach yields statistically significant improvements in at least one regime for each dataset.
Lastly, we conduct an ablation study that quantifies the impact on performance of: \emph{i)} MSB, \emph{ii)} the Learning to Reweight mechanism, and \emph{iii)} the degree of sparsity in the training dataset.

The main contributions of our paper can be hence summarized as follows:

\begin{itemize}

\item We introduce a robust semi-supervised framework for link polarity prediction.
Our model-agnostic framework dynamically learns the importance of each sample labeled using multiscale social balance theory.

\item We rigorously characterize the efficacy of our framework under conditions characterized by sparse labeling and significant noise, showcasing the robustness of our approach.

\item We validate the performance enhancements provided by our framework against other semi-supervised and robust learning approaches from the literature, quantifying the improvements in the link polarity prediction task.

\item Through detailed ablation studies, we assess the importance of
integrating multiscale social balance,
providing insights into the specific contributions and optimizations that are most effective for improving prediction accuracy.

\end{itemize}

The rest of the paper is organized as follows. Section~\ref{sec:related} reviews the literature most related to our work, highlighting the differences with our proposal. Sections~\ref{sec:problem} and~\ref{sec:method} formally set the context of our approach and its technical details. Section~\ref{sec:experiments} provides an experimental validation of the approach, and finally Section~\ref{sec:conclusions} discusses open problems and provides directions for further improvements.

%% file: related.tex
To the best of our knowledge, no prior work has addressed the problem of learning SGNNs under noisy and sparse dataset conditions: these two issues are related to Semi-Supervised Learning (SSL)~\cite{yang2022survey} and Robust Learning~\cite{song2022learning}. We now briefly describe how the current literature on these two topics relates to our problem setting.

\spara{SSL and robust learning.}
SSL leverages both labeled and unlabeled data to perform specific learning tasks.
However, current SSL approaches cannot be trivially extended to SGNNs for various reasons. Generative methods~\cite{kingma2014semi} are limited in adaptability since no generative models exist for signed graphs. Consistency regularization~\cite{berthelot2019mixmatch, sohn2020fixmatch} relies on alternative views of the input dataset but data augmentation methods for signed graphs are lacking. Graph-based methods~\cite{zhu2002learning, kipf2016semi, stretcu2019graph} can frame the link polarity prediction problem as a node classification problem on a line graph~\cite{cameron1991line}, but this approach faces significant scalability issues. By contrast, Pseudo-Labeling (PL) methods are inherently model-agnostic and can be easily adapted to any network~\cite{rizve2020defense, cascante2021curriculum}.
Regarding robust learning, RSGNN~\cite{zhang2023rsgnn} represents the state-of-the-art approach for learning SGNNs under noisy datasets.
Similar to our work, it proposes a model-agnostic training framework to improve SGNN training. However, there are three substantial differences with our proposal: \emph{i)} our framework can learn not only under noisy labels but also in the presence of an imbalance between labeled and unlabeled information, \emph{ii)} RSGNN cannot be applied to large graphs as it optimizes all the entries of the adjacency matrix, thus requiring at least $n^2$ parameters, and \emph{iii)} since RSGNN optimizes the adjacency matrix in the continuous space, it is not clear how to adapt neural modules that work with discrete structures such as graph attention.

\spara{Social balance in SGNN.}
As stated in the Introduction,  many SGNNs have been proposed that leverage the classic social balance theory~\cite{derr2018signed,huang2019signed,huang2021sdgnn,Shu21,kim2023trustsgcn}, at the microscale level, i.e., focusing on triads.
Building on the multiscale perspective, that considers both the triadic level and the subgroups/communities level, we propose a reweighting scheme to filter out cases where SB theory does not hold. A recent work~\cite{kim2023trustsgcn} explores a similar idea by refining the message-passing propagation of SGCN using a social balance reweighting classifier trained offline. Their contribution is orthogonal to ours, as they propose a new model, while we propose a training framework that enhances the performance of any given SGNN in sparse and noisy scenarios. Additionally, their reweighting is limited to microscale SB, undirected networks, and they learn the weights as a preprocessing step instead of optimizing them in an end-to-end fashion with the SGNN parameters.

%% file: problem.tex
We are given a directed signed graph $\mathcal{G} = (\mathcal{V}, \mathcal{E}^+, \mathcal{E}^-,\mathcal{E}^U)$ where $\mathcal{V} = (v_1, ..., v_n)$ represents the set of nodes, $\mathcal{E}^L = \mathcal{E}^+ \cup \mathcal{E}^-$ denotes the set of labeled edges, with  $\mathcal{E^+}$ and $\mathcal{E^-}$ denoting positive and negative edges respectively, while $\mathcal{E}^U$ denote a set of unlabeled edges, such that  $\mathcal{E}^L \cup \mathcal{E}^U \subseteq \mathcal{V}\times \mathcal{V}$ and
$\mathcal{E^+} \cap \mathcal{E^-} = \mathcal{E}^L\cap \mathcal{E}^U = \emptyset$.

Signed network representation learning aims to learn a node-embedding function $f_{\theta}: \mathcal{V} \rightarrow \mathbb{R}^d$ that maps each node $v_i$ to a point $z_i = f_\theta(v_i)$ in a $d$-dimensional latent space. The embedding of $\mathcal{V}$ in the latent space is denoted by $Z\in \mathbb{R}^{n\times d}$.
In this space, nodes linked by positive edges are closer together, while those linked by negative edges are distanced, reflecting their relationship polarities.

The objective is to use the embedding function $f_{\theta}$ to accurately predict the polarity of directed links between nodes.
In practice, we aim at devising a scoring function
$S_{\theta}: Z \times Z \rightarrow \{0, 1\}$, which exploits $f_{\theta}$ and predicts whether a prospective edge $(u,v) \in \mathcal{V}\times \mathcal{V} - \mathcal{E}^L$ is positive or negative based on the geometric proximity of node embeddings. We hence optimize $f_{\theta}$ to enhance the accuracy of $S_{\theta}$.
\begin{problem}
\begin{problemframe}
(\textbf{Semi-Supervised Signed Network Representation Learning}) Given a signed graph $\mathcal{G} = (\mathcal{V}, \mathcal{E}^+, \mathcal{E}^-,\mathcal{E}^U)$, embed each node $v \in \mathcal{V}$ into a $d$-dimensional embedding space through $f_\theta(v)\in\mathbb{R}^{d}$, thus obtaining $Z\in \mathbb{R}^{n \times d}$.
\end{problemframe}
\end{problem}

Moving a step forward, we consider that the edge labels that we know might be noisy. More specifically, what we observe is a noisy version of the true graph $G$, denoted $\Tilde{G} = (\mathcal{V}, \Tilde{\mathcal{E}}^+, \Tilde{\mathcal{E}}^-,\mathcal{E}^U)$, where $\Tilde{\mathcal{E}}^+$ (resp.  $\Tilde{\mathcal{E}}^-$) is a random subset of $\mathcal{E}^+ \cup \mathcal{E}^-$ selected according to some unknown noise distribution. Hence, this subset includes edges that are either correctly or incorrectly labeled. The research question here is whether we can devise an embedding that is robust to noise.
\begin{problem}
\begin{problemframe}
\textbf{(Robust Signed Network Representation Learning)}
Given a perturbed signed graph $\mathcal{\Tilde{G}}=(\mathcal{V}, \Tilde{\mathcal{E}}^+, \Tilde{\mathcal{E}}^-,\mathcal{E}^U)$ with noisy links, the goal is to learn an embedding function $f_{\theta}$
such that the embeddings derived from the noisy data approximate the true embeddings closely,  i.e. $\Tilde{Z} \approx Z$.
\end{problemframe}
\end{problem}

In the next section we shall devise a general framework for learning $f$ and $S$. Through the experiments in Section~\ref{sec:experiments} we will show that the framework is robust enough to cope with both problems. 

%% file: method.tex
\subsection{Preliminaries}
First, we model $f_\theta$ using an SGNN model that projects each node $v \in \mathcal{V}$ into a $d$-dimensional embedding space.
Each edge $\ell = (u, v)$ can possibly have a sign $\sigma(l) \in \{0, 1\}$, where without loss of generality 1 is used for positive polarity, and 0 for negative polarity. To alleviate the notation we will sometimes refer to $f_{\theta}$ as $f$.
Given an SGNN model $f$, we aim to predict whether a link $\ell=(u,v)$ polarity is positive or negative.
To achieve this, we devise a logistic probability score $s_\ell = (1 + \exp(-a))^{-1}$,
where $a=-\psi \cdot Z_\ell$, $Z_{\ell} = Z_u || Z_v$ represents the concatenated node embeddings of $u$ and $v$, and $\psi$ is a learnable parameter vector. This probability is then used to compute the binary cross-entropy loss over the sign of the link:
\begin{equation} \label{eq:loss-sign}
L_{\text{sign}}(\Theta, \ell) = (\sigma(\ell)-1)\log(1-s_{\ell}) -\sigma(\ell)\log(s_{\ell}),
\end{equation}
where $\Theta = \{\theta, \psi\}$ is the complete set of learnable parameters: $\theta$ correspond to the SGNN parameters, and $\psi$ to the final linear layer.
We also consider a task-specific loss function $L_{\text{task}}(\theta, \mathcal{G})$, designed to impose additional constraints on the embedding based on the structural properties of interest in the graph $\mathcal{G}$. For example, $L_{\text{task}}$ can encode constraints related to the SGNN structure of $f$ or ensure that embeddings with the same sign are geometrically close.

\subsection{Framework}
We next introduce our model-agnostic framework to learn $f$ from $\mathcal{G} = (\mathcal{V}, \mathcal{E}^+, \mathcal{E}^-,\mathcal{E}^U)$, independently of the underlying SGNN model adopted. We leverage Multiscale Social Balance (MSB) prior knowledge together with the unsupervised information to improve the performance of $f$, especially when dealing with sparsely-labeled and noisy datasets.

\spara{MSB Labeling.}
As shown in Fig.~\ref{fig:social-balance-explain},
we can combine prior knowledge about SB along with the set of unlabeled edges $\mathcal{E}^U$ to derive a set of edges $\mathcal{E}_{\text{SB}} \subseteq \mathcal{E}^U$ with associated labels.
For the microscale level,
we enumerate all transitive triads $\Delta \subset \mathcal{E} \times  \mathcal{E} \times  \mathcal{E}$, we select only those triads with exactly one unlabeled edge, denoted as $\Delta^{1U}$.
For each triad in $\Delta^{1U}$,
we add the unlabeled edge $(u,v)$ to $\mathcal{E}_{\text{SB}}$ and the sign $\sigma(u,v)$ to $\mathcal{S}_{\text{SB}}$ derived through the microscale SB theory, i.e., so that the product of signs of the triad is positive  (Fig.~\ref{fig:social-balance-explain}a).

For the mesoscale level,
we assume to have at disposal a partitioning of $\mathcal{V}$ in $k$ communities, i.e, $\mathcal{C} = \{c_1, ..., c_k\}$ such that $\cup_{i = 1}^k c_i = \mathcal{V}$ and $c_i \cap c_j = \emptyset$ for $i, j \in [k]$, $i \neq j$. This can be computed using any state of the art graph partitioning or community detection method.
For each unlabeled edge $(u,v) \in \mathcal{E}^U$, we add $(u,v)$ to $\mathcal{E}_{\text{SB}}$ and the sign $\sigma(u,v)$ to $\mathcal{S}_{\text{SB}}$ such that the mesoscale SB theory is satisfied $c$, i.e, edges crossing communities are negative, while edges within communities are positive (Fig.~\ref{fig:social-balance-explain}b).

\spara{Learning to reweight unknown labels.}
A problem that we need to address is the reliability of the sign $\sigma(\ell)$ associated with each edge $\ell \in \mathcal{E}_{\text{SB}}$. The presence of noisy edges in $\mathcal{E}^L$ makes the SB assignment potentially unreliable. This is especially true at a mesoscale level, where labeling is a result of a community-detection algorithm that could be potentially biased by the underlying heuristics and hence introduce some noise.
When learning $f$, we consequently need to deal with potential uncertainty.

We approach this problem by resorting to a Social Balance weight model $w:\mathcal{E}_{\text{SB}} \rightarrow [0,1]$,  which can potentially be learned alongside the embedding model $f$. Since Multiscale Social Balance (SB) does not always hold, the idea is to learn how to reweight the examples in $\mathcal{E}_{\text{SB}}$. In practice, the learning of $\Theta$ can benefit from considering unknown edges for which social-balance theory applies. However, the contribution of these edges needs to be weighted, with some edges more reliable than others.
Thus, the problem we want to approach can be stated as follows. Provided that the optimal values of $\Theta$ can be considered a function of $w$, i.e.,
\begin{equation}\label{eq:optim}
\Theta^*(w) = \arg \min_{\Theta}\sum_{\ell_i\in \mathcal{E}^L\cup \mathcal{E}_{\text{SB}}} \hat{w}_iL_{\text{sign}} (\Theta, \ell_i),
    \end{equation}
where
$$
\hat{w}_i = \left\{\begin{array}{lr}
    1 & \mbox{if }\ell_i \in \mathcal{E}^L \\
    w_i & \mbox{otherwise}, \\
\end{array}\right.
$$
we aim to learn the weights $w$ that are most appropriate to improve the model $\Theta^*(w)$.

The approach we adopt is inspired by~\cite{ren2018learning}.
We assume that the number of edges in $\mathcal{E}^{\text{U}}$ is orders of magnitude greater than the number of edges in $\mathcal{E}^L$. We can hence expect that the optimization described in Eq.~\ref{eq:optim} is mainly influenced by the edges in $\mathcal{E}_{\text{SB}}$. This assumption is realistic in the scenarios described in the introduction, where the amount of labeled data is scarce. Thus, we can devise the optimization approach in two steps:
\begin{itemize}
    \item We find an approximation of the optimal value for $\Theta(w)$ by only considering edges in $\mathcal{E}_{\text{SB}}$.
    \item We treat $w$ as a set of hyperparameters, whose optimization involves checking whether the optimal solution  $\Theta^*(w)$ is also optimal for the separate set of edges  $\mathcal{E}^L$, which serves as a validation set.
\end{itemize}
As described in~\cite{ren2018learning}, for this we can devise an alternated online updating procedure with the objective of progressively learning small weights for examples where Social Balance does not stand true and higher weights viceversa, when optimizing the model parameters.

Formally, for a batch $\mathcal{B}$ of edges, we can devise a loss function
\begin{equation}\label{eq:weighted_loss}
l(w,\Theta;\mathcal{B}) = \sum_{\ell_i\in \mathcal{B}} \hat{w}_iL_{\text{sign}} (\Theta,\ell_i).
\end{equation}
Let us consider now two subsets $\mathcal{B}_{\text{clean}}\subseteq\mathcal{E}^L$ and $\mathcal{B}_{\text{SB}}\subseteq\mathcal{E}_{\text{SB}}$ of edges. Then, for a given model instance $\Theta$, we can get a differentiable improved version  $\tilde{\Theta}$ using $\mathcal{B}_{\text{SB}}$ by updating $\Theta$ in the direction of its gradient. For example, we can define it as:
$$
\tilde{\Theta} =  \Theta - \alpha \nabla_\Theta l(w,\Theta;\mathcal{B}_{\text{SB}})
$$
This new value can be exploited to get a update of $w$ that improves the response on $\mathcal{B}_{\text{clean}}$. Let
$$
    \epsilon_i  = \left\{\begin{array}{ll}
        w_i -\eta \frac{\partial}{\partial {w_i}} l(w_i,\tilde{\Theta};\mathcal{B}_{\text{clean}}) & \mbox{if } \ell_i \in \mathcal{B}_{\text{SB}}\\
        w_i & \mbox{otherwise.}
    \end{array} \right.
$$
Then, we can update $w$ by averaging over $\epsilon$:
$$
w_i = \frac{\max(0,\epsilon_i)}{\sum_{j\in \mathcal{E}_{\text{SB}}} \max(0,\epsilon_j)}
$$

\spara{Learning procedure.}
Algorithm~\ref{alg:lrw_sb} embodies the outlined approach to obtain the necessary weights.
We initialize the weights $\epsilon$ with random values drawn uniformly from the interval $(0,1)$. Subsequently, we compute the sign loss on batch $\mathcal{B}_{\text{SB}}$ weighting each edge by the corresponding $\epsilon_i$, using the node embeddings as per the current model $f_{\theta^{(t)}}$ (lines 1-2). We then progress to construct a meta-model $f_{\hat{\theta}^{(t)}}$ by updating the $\theta^{(t)}$ parameters (lines 3-4). Leveraging the meta-model, we define a novel loss function $l_{\text{clean}}$, grounded on the embeddings of the batch $\mathcal{B}^{\text{clean}}$ computed via $f_{\hat{\theta}^{(t)}}$.
 Subsequent to computing the gradients concerning $\epsilon$, we adjust the uniform weights utilizing $\nabla \epsilon$ (lines 6-7). Finally, to mitigate negative weights, we subject the outcomes to a Rectified Linear Unit (ReLU) transformation and normalize the resultant weights $w$ to lie within the interval $[0,1]$ (line 8). The \texttt{BackwardAD} component symbolizes any automated differentiation mechanism facilitating the computation of gradients.
\begin{algorithm}
\caption{$\texttt{ReWeight}(\mathcal{V},\mathcal{B}_{\text{clean}},\mathcal{B}_{\text{SB}},\Theta$)}
\label{alg:lrw_sb}
\KwData{Node set $\mathcal{V}$,
batches $\mathcal{B}_{\text{clean}}, \mathcal{B}_{\text{SB}}$  of clean and multiscale-SB labeled edges,
model parameters $\Theta$ and underlying embedding model $f_{\theta}$,
}
\KwResult{weight model $w: \mathcal{E}_{\text{SB}} \rightarrow [0,1]$}

$\epsilon \leftarrow \texttt{Uniform}(0,1)$; \\
$l_{\text{SB}} \leftarrow l(\epsilon,\Theta,\mathcal{B}_{\text{SB}})$ \tcp*{eq.~\ref{eq:weighted_loss}}
\tcc{Compute meta-model}
$\nabla\Theta \leftarrow \texttt{BackwardAD}(l_{\text{SB}}, \Theta)$ \\
$\hat{\Theta} \leftarrow \Theta^{(t)} - \alpha\nabla\Theta$ \\
$l_{\text{clean}} \leftarrow \frac{1}{|\mathcal{B}^{\text{clean}}|}\sum_{\ell_i \in \mathcal{B}^{\text{clean}}}  L_{\text{sign}}(\hat{\Theta} ,\ell_i)$ \\
\tcc{Compute weights using meta-model}
$\nabla_{\epsilon} \leftarrow \texttt{BackwardAD}(l_{\text{clean}}, \epsilon)$ \\
$\Tilde{w} \leftarrow max(0, \epsilon - \nabla_{\epsilon})$\tcp*{Avoid negative weights}
$w \leftarrow \frac{\Tilde{w}}{\sum_{\ell_i\in\mathcal{B}_{\text{SB}}} \Tilde{w}_i}$ \\
\end{algorithm}

This reweighting procedure is employed within a global learning framework that we call \ourProcedure{}, depicted in Algorithm~\ref{alg:signed_gnn_ss}. The process starts by selecting edges with signs conforming to social balance principles (as previously described in the "Multiscale SB Labeling" paragraph), after which it enters the main learning loop. This loop comprises two key steps: initially, we compute the weights utilizing Algorithm~\ref{alg:lrw_sb} (line 5). Subsequently, these weights are used to blend contributions from both the task-specific loss $L_{\text{task}}$ and the weighted sign loss $L_{\text{sign}}$. Finally, the model undergoes updating by backpropagating the gradient of the combined loss with respect to the current model parameters $\Theta^{(t)}$ (lines 6-9).

\begin{algorithm}
\caption{\ourProcedure{}}
\label{alg:signed_gnn_ss}
\KwData{Signed Graph $\mathcal{G}=(\mathcal{V}, \mathcal{E}^+, \mathcal{E}^-, \mathcal{E}^U)$,
signed GNN model $f$,
loss function $L_{\text{task}}$
}
\KwResult{Model parameters $\Theta$ and underlying (trained) embedding model $f_{\theta}$}

Extract edges and signs $\{\mathcal{E}_{\text{SB}}, \mathcal{S}_{\text{SB}}\}$ from $\mathcal{E}^U$ using Multiscale SB \\
Initialize $\Theta^{(0)}$\\
\While(\tcp*[f]{iterate over $t$}){training does not converge}{
    Sample $\mathcal{B}_{\text{clean}}\subseteq \mathcal{E}^+ \cup \mathcal{E}^-$ and $\mathcal{B}_{\text{SB}}\subseteq\mathcal{E}_{\text{SB}}$\\
    $w \leftarrow \texttt{ReWeight}(\mathcal{V},\mathcal{B}_{\text{clean}},\mathcal{B}_{\text{SB}},\Theta^{(t)})$ \tcp*{learn weights}
        $l_{\text{task}}=L_{\text{task}}(\theta^{(t)},\mathcal{G})$ \tcp*{Compute task loss }
        $l_{\text{SB}} = l(w,\Theta^{(t)},\mathcal{B}_{\text{SB}})$ \tcp*{eq.~\ref{eq:weighted_loss}}
        $l_{\text{tot}} = l_{\text{task}} + l_{\text{SB}}$ \tcp*{Compute the total loss}
        $\Theta^{(t+1)} \leftarrow \texttt{BackPropagate}(l_{\text{tot}})$ \tcp*{Update model}
    }
\end{algorithm}
\vspace{-6mm}
\spara{Training \& Inference time.}
The implementation of the Learning to Reweight Social Balance methodology, as illustrated in Algorithm~\ref{alg:lrw_sb}, introduces a constant additional load on the training time. Algorithm~\ref{alg:signed_gnn_ss} requires performing two forward and two backward operations, resulting in twice the overhead compared to the standard training procedure of a SGNN model. However, it is worth noting that this overhead only affects training time, while inference time remains solely based on the computation of the SGNN model.

\spara{Practical details.}
Algorithm~\ref{alg:signed_gnn_ss} operates on batches of edges to compute contributions to the losses. Regarding $\mathcal{B}_{\text{clean}}$,
our experiments revealed a useful rule-of-thumb: randomly sample approximately half of the edges  in $\mathcal{E}^L$ to build the clean set. By contrast, $\mathcal{B}_{\text{SB}}$ is sampled to include a larger number of edges compared to $\mathcal{B}_{\text{clean}}$. In our experiments, this ratio typically stands at 6:1.

\subsection{Convergence}
\label{subsec:convergence}
The main assumption of the Learning to Reweight procedure~\cite{ren2018learning} used in our framework %
relies on the Lipschitz smoothness of the loss function, which ensures that reweighting converges to the critical loss value on the clean dataset.
The Lipschitz smoothness of a scalar-valued function $f:\mathbb{R}^d \rightarrow \mathbb{R}$ can be defined, in its infinitesimal form, by the condition:
\[
||\nabla_{\vec{x}} f(\vec{x})|| \leq K
\]
where $K$ is a constant.

In our study, we utilize the Binary Cross-Entropy (BCE) loss, described by Eq.~\ref{eq:loss-sign}. The gradient of this loss function is a two-dimensional vector:
\begin{equation}
\begin{split}
\nabla L_{\text{sign}}(\Theta, \ell) & =  \nabla_{s_\ell} L_{\text{sign}} \cdot s_\ell'\cdot \begin{pmatrix}
     \nabla_{\psi} a  \\
      \nabla_{Z_\ell} a \cdot \nabla_{\theta} Z_\ell
\end{pmatrix} \\
 & =  \nabla_{s_\ell} L_{\text{sign}} \cdot s_\ell'\cdot \begin{pmatrix}
     Z_\ell  \\
      \psi \cdot \nabla_{\theta} Z_\ell
\end{pmatrix}
    \end{split}
\end{equation}
where the first component corresponds to the update of the parameters of the final layer ($\psi$) and the second component to the update of the parameters of the GNN ($\theta$).
By inspecting its components, we can observe the following. The first term is the derivative of the BCE, and it can be expressed as
    $$
    \nabla_{s_\ell} L_{\text{sign}}= \frac{1-\sigma(\ell)}{1-s_{\ell}} - \frac{\sigma(\ell)}{s_{\ell}}.
    $$
    To prevent the singularities that occur as $s_{\ell}$ approaches 0 or 1, we clamp the log probabilities in the computation of the BCE loss. This adjustment ensures that the gradients of the loss function remain finite and well-defined, as also recommended by the PyTorch documentation\footnote{\url{https://pytorch.org/docs/stable/generated/torch.nn.BCELoss.html}}.
The second term represents the derivative of the logistic function, which is limited within the interval $[0, 1/4]$.
For the third term, we observe that the amplitudes of both the $Z_l$ and $\psi$ components can be controlled through regularization. The last term, $\nabla_{\theta} Z_\ell$, depends on the structure of the underlying GNN. Typically, the use of linear transformations along with activation functions that have bounded gradients—common in most GNN implementations—ensures that this component remains bounded as well.
Notice that similar considerations should be made for the $L_{\text{task}}$ component, which we consider parametric in our framework.

%% file: experiments.tex
This experimental section examines whether our framework effectively predicts link polarities within sparse and noisy datasets. We aim to predict the signs of links not observed during the training phase.
We design the experiments to address the following research questions:

\begin{itemize}
    \item \textbf{RQ1}: Can our semi-supervised framework significantly improve the performance of an SGNN in regimes with sparse labels and noise?
    \item \textbf{RQ2}: How does the performance of our framework compare to other semi-supervised and robust learning baselines?
    \item \textbf{RQ3}: To what extent do the Micro and Mesoscale Social Balance components contribute to the informativeness of our framework?
    \item \textbf{RQ4}: What is the importance of learning to reweight MSB within our framework?
    \item \textbf{RQ5}: What are the implications of leveraging varying degrees of unlabeled-to-labeled signal within our framework?
\end{itemize}

\subsection{Experimental settings}
\label{subsec:exp-settings}
\spara{Datasets.}
We conduct our experiments using four distinct signed graph datasets sourced from the realm of social networks:
\begin{itemize}
    \item Bitcoin: Alpha and OTC denote two online exchanges facilitating bitcoin transactions. Within these platforms, users have the capability to rate others using a trust metric spanning from -10 to 10. A rating of -10 signifies individuals flagged as potential fraudsters, while a rating of +10 indicates complete trustworthiness. For our purposes, we binarize the edges by considering negative all the edges with negative rating, and accordingly as positive all the positive ones. 
    \item Wiki: This dataset describes voting dynamics pertinent to the election of Wikipedia admins. Positive edges signify supporting votes, whereas negative edges denote opposing votes.
    \item Slashdot: This dataset is derived from a technology-news platform and comprises connections between users, representing both affiliations and antagonisms within the community.
\end{itemize}
The summary statistics for the four datasets are provided in Tab.~\ref{tab:data-stats}. 
For a detailed description, we point the reader to~\cite{he2022pytorch}.
\begin{table}
\caption{Dataset statistics including the empirical ratio of balanced triads \% MicroSB and balanced intra and inter-community edges \% MesoSB.}
\fontsize{6.85}{7}\selectfont %
\centering
\begin{tabular}{lccccc}
     \toprule
     Dataset & $|\mathcal{V}|$ & $|\mathcal{E}^+|$ & $|\mathcal{E}^-|$ & \% MicroSB & \% MesoSB \\
     \midrule
     Bitcoin-Alpha & 3,783 & 22,650 & 1,536 & 88 \% & 62 \% \\
     Bitcoin-OTC & 5,881 & 32,029 & 3,563 & 89 \% & 63 \% \\
     Wiki & 11,259 & 138,813 & 39,283 & 75 \% & 59 \% \\
     Slashdot & 82,140 & 425,072 & 124,130 & 91 \% & 53 \% \\
     \bottomrule
\end{tabular}

\label{tab:data-stats}
\end{table}
The table presents standard metrics such as the number of nodes and the distribution of positive and negative edges. Additionally, it indicates the extent to which both micro and mesoscale social balance theories hold within these networks. For the microscale, we show the percentage of consistent triads. By contrast, the mesoscale measures the number of edges within or across communities that do not violate consistency.

We also measure how micro and mesoscale social balance theories compare to expectations derived from a null model. 
This null model, which we developed to evaluate the statistical significance of our findings, randomly swaps two edges of the same sign while preventing the formation of multiple edges between the same node pairs, regardless of the sign.
The model maintains the original in/out-degree and positive/negative-degree sequences of the graph.
Following the methodology described by \citet{uzzi2013atypical}, we perform 50 samples by performing $|\mathcal{E}| \log{|\mathcal{E}|}$ swaps for each sample. This process is critical for comparing the empirical frequencies of balanced patterns observed in our datasets with those expected under random conditions. Tab.~\ref{tab:null-model} reports, for each balanced triad type, the empirical frequency relative to each dataset, and the z-score and standard deviation relative to the null model. We can observe that all the z-scores exhibit high values: a clear sign that that each dataset embeds relevant structures relative to the signs. We further notice that `+++' triads represent an anomalous vast majority in the Wiki and Slashdot datasets, and in general for these two dataset the deviance from the null model is extremely significant.  

\begin{table}
\caption{Z-scores of the empirical frequency of each balanced triad type compared to their frequency in the null model.}
\fontsize{6.85}{7}\selectfont %
\centering
\begin{tabular}{lcccc}
\toprule
Dataset & Triad  & Emp Frequency & z-score & Std Dev \\
\midrule
Bitcoin-Alpha & + + +  & 74,632 & 19.26 & 1003.25 \\
Bitcoin-Alpha & + - -  & 2,492 & 11.04 & 96.24 \\
Bitcoin-Alpha & - + -  & 951 & 11.34 & 40.05 \\
Bitcoin-Alpha & - - +  & 550 & 9.59 & 27.93 \\
\midrule
Bitcoin-OTC & + + +  & 103,194 & 15.68 & 1360.81 \\
Bitcoin-OTC & + - -  & 4,391 & 17.56 & 116.01 \\
Bitcoin-OTC & - + -  & 3,921 & 30.22 & 91.57 \\
Bitcoin-OTC & - - +  & 1,478 & 16.73 & 46.52 \\
\midrule
Wiki & + + +  & 1,013,594 & 298.86 & 2030.64 \\
Wiki & + - -  & 75,692 & 148.24 & 286.07 \\
Wiki & - + -  & 53,699 & 94.48 & 339.59 \\
Wiki & - - +  & 21,719 & 36.53 & 236.85 \\
\midrule
Slashdot & + + +  & 1,034,150 & 283.21 & 2671.24 \\
Slashdot & + - -  & 25,385 & 72.34 & 208.26 \\
Slashdot & - + -  & 64,934 & 90.31 & 403.06 \\
Slashdot & - - +  & 18,176 & 47.14 & 198.31 \\
\bottomrule
\end{tabular}
\label{tab:null-model}
\end{table}

We perform the same analysis at the mesoscale level. The results shown in Tab.~\ref{tab:null-model-meso} confirm that mesoscale SB also exhibits high z-scores and standard deviations, signifying that these patterns are not due to chance. Interestingly, for Bitcoin-Alpha and Bitcoin-OTC, the z-scores for mesoscale SB are higher than those for microscale SB; conversely, for Wiki and Slashdot, the opposite trend is observed.

\begin{table}
\caption{Z-scores of the empirical frequency of the mesoscale social balance compared to its frequency in the null model.}
\centering
\fontsize{6.85}{7}\selectfont %
\begin{tabular}{lccc}
\toprule
Dataset & Emp Frequency & z-score & Std Dev \\
\midrule
Bitcoin-Alpha & 14,787 & 34.55 & 199.84 \\
Bitcoin-OTC & 23,274 & 40.19 & 282.36 \\
Wiki & 116,247 & 30.83 & 1702.46 \\
Slashdot & 291,440 & 12.11 & 7007.47 \\
\bottomrule
\end{tabular}
\label{tab:null-model-meso}
\end{table}
\vspace{-0.15cm}
\spara{Baselines.}
In this paragraph, we introduce the baseline methods used for comparison with our proposed approach. We aim to compare our proposal with semi-supervised and robust learning frameworks capable of training an SGNN model under sparse and noisy dataset conditions.
For the semi-supervised baselines, we adopt a general Pseudo-Labeling (PL) approach, which involves training a base model $f$ using the labeled edges and then repeating the following process for $T$ iterations: 
$i)$ Select $K$ edges from $\mathcal{E}^U$; 
$ii)$ Assign a pseudo-label to each of the $K$ edges using $f$;
$iii)$ Merge the pseudo-labeled edges with the supervised edges set;
$iv)$ Retrain $f$ on the augmented dataset from point $iii$. 
The model with the highest Macro-F1 on the validation set across all the $T$ iterations is then returned.
We implement three different versions of this general PL approach by refining the selection process at point $i)$:
\begin{itemize}
    \item PL (All): Select all examples from $\mathcal{E}^U$, limiting the procedure to one iteration.
    \item PL (Random): Randomly sample $K$ examples from $\mathcal{E}^U$.
    \item PL (Uncertainty): Select the $K$ examples for which $f$ is most certain on its labeling. To address the data imbalance, we select the $\frac{K}{2}$ edges with the lowest prediction score and the $\frac{K}{2}$ with a higher score.
\end{itemize}
The last baseline implements suggestions from~\cite{rizve2020defense} to consider model uncertainty when selecting instances to pseudo-label. We restart the model parameters before retraining (step $iv$) to incorporate findings from~\cite{cascante2021curriculum}. 
We set $K$ equal to $50$ and $T$ equal to $10$. Altering these two hyperparameters does not significantly affect the results.

Regarding robust learning competitors, we have selected the recent RSGNN approach proposed by~\citet{zhang2023rsgnn}, because it addresses the issue of Robust Link Polarity prediction. RSGNN is based on jointly learning the SGNN parameters and a denoised adjacency matrix to ensure robustness in a noisy data scenario, thus forming a suitable baseline for comparison.

\spara{Implementation details.}
We utilize the PyTorch Geometric Signed Directed library to implement the data loading and modeling components~\cite{he2022pytorch}. Given that Link Polarity Prediction is a binary classification task, we employ Accuracy and Macro-F1 metrics for evaluation, with the latter addressing label imbalance.
We allocate 5\% of the edges for validation, 5\% for testing. To create a sparse scenario, we further partition 75\% of the remaining edges into an unlabeled set, i.e., $\mathcal{E}^U$. The remaining 25\% is used for the labeled part, i.e., the labeled edges $\mathcal{E}^L$.
The splits for training-validation-test and unlabeled-labeled are performed twenty times to report the average over the runs and to assess statistical significance using a paired t-test.
Since we do not have the user-level communities $\mathcal{C}$ at our disposal, we derive them as a preprocessing step using the Louvain algorithm~\cite{blondel2008fast} on the unsigned version of $\mathcal{G}$. For the node attributes, we generate feature vectors using the spectral decomposition of the adjacency matrix, as commonly done in the literature~\cite{huang2021sdgnn}. 
We set the embedding dimension to 64 and use the Adam Optimizer with a learning rate and weight decay of $1e^{-3}$. Training is conducted for 1000 epochs, evaluating the Macro-F1 on the validation set every 25 epochs, with early stopping implemented after 10 epochs of no improvement. 
We select SDGNN~\cite{huang2021sdgnn} as the base model, as it supports directed signed graphs. The SDGNN model is configured with 2 convolutional layers, and we use the SDGNN loss (with default hyper-parameters) as the "task" loss $L_{\text{task}}$.
Our experiments~\footnote{Code repository: \href{\repo}{\repo}} are conducted on a DGX Server equipped with 4 NVIDIA Tesla V100 GPU (32GB) and CUDA Version 12.2. 

\subsection{Link Polarity prediction}
\spara{(RQ1) The proposed framework enhances the performance of an SGNN in sparse and noisy regimes.}
\begin{table}
    \centering
    \setlength{\tabcolsep}{3pt} %
    \renewcommand{\arraystretch}{1.1} %
    \fontsize{6.85}{7}\selectfont %
    \begin{tabular}{cccccc}
         \toprule
         \multirow{2}{*}{Dataset} & \multirow{2}{*}{Perturb (\%)}  & \multicolumn{2}{c}{SDGNN} & \multicolumn{2}{c}{SDGNN+\ourProcedure} \\
         \cmidrule(lr){3-4} \cmidrule(lr){5-6}
         {} & {} & Accuracy & Macro-F1 & Accuracy & Macro-F1 \\
         \midrule
         \multirow{2}{*}{Bitcoin-Alpha} & 0 & 0.9100 & 0.6734 & $\textbf{0.9322}^{***}$ & $\textbf{0.6846}^{*}$ \\
         {} & 10 & 0.7792 & 0.5405 & $\textbf{0.8368}^{***}$ & $\textbf{0.5897}^{***}$ \\
         {} & 20 & 0.7550 & 0.5037 & $\textbf{0.8512}^{***}$ &  $\textbf{0.5614}^{***}$ \\
         \midrule
         \multirow{2}{*}{Bitcoin-OTC} & 0 & 0.8950 & 0.7210 & $\textbf{0.9054}$ & $\textbf{0.7438}^{***}$ \\
         {} & 10 & 0.7818 & 0.6147 & $\textbf{0.8487}^{***}$ & $\textbf{0.6672}^{***}$ \\
         {} & 20 & 0.7198 & 0.5549 & $\textbf{0.8469}^{***}$ & $\textbf{0.6337}^{***}$ \\
         \midrule
         \multirow{2}{*}{Wiki} & 0 & 0.8242 & 0.7591 & $\textbf{0.8354}^{***}$ & $\textbf{0.7634}^{***}$ \\
         {} & 10 & 0.7856 & 0.7197 & $\textbf{0.8030}^{***}$ & $\textbf{0.7297}^{***}$ \\
         {} & 20 & 0.7441 & 0.6851 & $\textbf{0.7821}^{***}$ & $\textbf{0.7054}^{***}$ \\
         \midrule
         \multirow{2}{*}{Slashdot} & 0 & 0.8095 & 0.7439 & $\textbf{0.8220}^{**}$ & $\textbf{0.7520}^{**}$ \\
         {} & 10 & 0.7522 & 0.6905 & $\textbf{0.7634}^{*}$ & $\textbf{0.6963}^*$ \\
         {} & 20 & 0.7327 & 0.6710 & $\textbf{0.7466}^{**}$ & $\textbf{0.6763}^{*}$ \\
         \bottomrule
    \end{tabular}
    \vspace{1mm}
    \caption{Accuracy and Macro-F1 for link polarity prediction on four benchmark datasets between SDGNN and SDGNN+\ourProcedure. Bold values denote the best model. Each value is the average across 20 different initial data splits. Asterisks denote the significance level according to a two-sided paired t-test compared to the second-best method (* if $<0.05$, ** if $<0.01$, *** if $<0.001$).}
    \label{tab:sdgnn-macrof1-main-results}
\end{table}
\begin{table*}[t]
    \centering
    \setlength{\tabcolsep}{3pt} %
    \renewcommand{\arraystretch}{1.1} %
    \fontsize{6.85}{7}\selectfont %
    \begin{tabular}{cccccccc}
         \toprule
         \multirow{2}{*}{Dataset} & \multirow{2}{*}{Perturb (\%)}  & Base Signed GNN & Robust Signed GNN & \multicolumn{3}{c}{Semi-Supervised Strategies} & Ours \\
         \cmidrule(lr){5-7}
         {} & {} & SDGNN & RSGNN & SDGNN + PL (All) & SDGNN + PL (Random) & SDGNN + PL (Uncertainty) & SDGNN + \ourProcedure{} \\
         \midrule
         \multirow{2}{*}{Bitcoin-Alpha} & 0 & 0.6734 & 0.6422 & 0.6726 & 0.6729 & $\underline{0.6757}$ & \textbf{0.6846} \\
         {} & 10 & 0.5405 & 0.5406 & 0.5359 & $\underline{0.5492}$ & 0.5470 & $\textbf{0.5897}^{***}$ \\
         {} & 20 & $\underline{0.5037}$ & 0.5024 & 0.5013 & 0.5015 & 0.5021 & $\textbf{0.5614}^{***}$ \\
         \midrule
         \multirow{2}{*}{Bitcoin-OTC} & 0 & 0.7210 & 0.6789 & 0.7319 & 0.7334 & $\underline{0.7337}$ & $\textbf{0.7438}^{***}$ \\
         {} & 10 & 0.6147 & 0.6058 & 0.6106 & 0.6122 & $\underline{0.6194}$ & $\textbf{0.6672}^{***}$ \\
         {} & 20 & 0.5549 & $\underline{0.5615}$ & 0.5474 & 0.5528 & 0.5581 & $\textbf{0.6337}^{***}$ \\
         \midrule
         \multirow{2}{*}{Wiki} & 0 & 0.7591 & 0.7200 & 0.7569 & $\underline{0.7592}$ & 0.7586 & $\textbf{0.7634}^{***}$ \\
         {} & 10 & $\underline{0.7197}$ & 0.6915 & 0.7103 & 0.7145 & 0.7158 & $\textbf{0.7297}^{***}$ \\
        {} & 20 & 0.6851 & 0.6781 & 0.6859 & 0.6856 & $\underline{0.6910}$ & $\textbf{0.7054}^{***}$ \\
         \midrule
         \multirow{2}{*}{Slashdot} & 0 & 0.7439 & NA & 0.7450 & $\underline{0.7497}$ & 0.7492 & $\textbf{0.7520}$ \\
         {} & 10 & 0.6905 & NA & 0.6893 & 0.6925 & $\underline{0.6953}$ & $\textbf{0.6963}$ \\
         {} & 20 & 0.6710 & NA & $\underline{0.6715}$ & 0.6703 & 0.6708 & $\textbf{0.6763}^{*}$ \\
         \bottomrule
    \end{tabular}
    \vspace{1mm}
    \caption{Macro-F1 for link polarity prediction on four benchmark datasets. Bold values denote the best model, and underlined values indicate the second best. Each value is the average across 20 different initial data splits. Asterisks denote the significance level according to a two-sided paired t-test compared to the second-best method (* if $<0.05$, ** if $<0.01$, *** if $<0.001$).}
    \label{tab:macrof1-main-results}
\end{table*}
We aim to determine whether our devised procedure can improve the performance of an SGNN model when learning from datasets that are sparse and contain noise in the labels (i.e., edge signs). 
To this end, we produce sparse versions of the datasets as described in Sec.~\ref{subsec:exp-settings} and train a base SGNN model, SDGNN~\cite{huang2021sdgnn}, with and without our procedure described in Alg.~\ref{alg:signed_gnn_ss}. In practice, we compare \ourProcedure{} with a basic SDGNN trained by only using labeled data. 
We also investigate the performance as the input signed graph contains increasing amounts of noise. 
The results are presented in Tab.~\ref{tab:sdgnn-macrof1-main-results}.

We observe that our \ourProcedure{} consistently produces statistically significant improvements in both Accuracy and Macro-F1 across all datasets, except for Accuracy on Bitcoin-OTC. The improvements are significant even when the datasets do not contain noise, highlighting the capability of our training procedure to extract important knowledge from the unlabeled edges using multiscale SB. Specifically, we see an Accuracy gain from 1.2\% in the case of Bitcoin-OTC without noise to 17.6\% on Bitcoin-OTC with 20\% perturbed labels. Averaged across all datasets, there is a 1.8\%, 4.9\%, and 9.3\% increase in Accuracy when 0\%, 10\%, and 20\% of edges are perturbed, respectively. 
Regarding the Macro-F1 metric, the minimum gain is 0.6\% on Wiki with 0\% noise, up to a maximum of 14.2\% for Bitcoin-OTC with 20\% noise. When we analyze the average across all datasets, we observe a 1.6\%, 4.9\%, and 7.3\% improvement in Macro-F1 when 0\%, 10\%, and 20\% of edges are perturbed, respectively. 

Notably, we observe consistently lower Macro-F1 improvements for the Wiki and Slashdot datasets, potentially correlated with high z-scores and the overrepresentation of specific triads, such as `+++' (see Tab.~\ref{tab:null-model}), which hinders the learning of the negative class. This suggests an underlying complexity within these networks, possibly involving dense balanced communities that play a prominent role in the embedding, disrupted by the randomization in the null model.

To summarize, our experiments demonstrate that \ourProcedure{} can significantly enhance the performance of a given SGNN, in this case, SDGNN~\cite{huang2021sdgnn}, with performance gains becoming more prominent as the noise in the dataset increases.

\spara{(RQ2) The proposed framework outperforms alternatives in the benchmark.}
In this analysis, we compare our \ourProcedure{} with other methods from the literature. We implement three semi-supervised learning frameworks based on Pseudo-Labeling: PL (All), PL (Random), PL (Uncertainty), and RSGNN, the state-of-the-art method for robust link polarity prediction. We report the results for the Macro-F1 metric in Tab.~\ref{tab:macrof1-main-results}. 
We limit the analysis to Macro-F1 for brevity and because it is the most challenging metric to optimize, given that it accounts for the data imbalance of negative polarities.

Our results show that \ourProcedure{} systematically achieves the best performance, obtaining statistically significant improvements in all cases where there is some noise in the data, with the exception of Slashdot with 10\% noise. On sparse datasets with no noise, it achieves statistically significant better performance on two out of four datasets. Averaged across all datasets, \ourProcedure{} shows improvements of 0.9\%, 4.1\%, and 6.9\% against the best of the competitors when the dataset has 0\%, 10\%, and 20\% noise, respectively.

Notably, we have no results for RSGNN on Slashdot. 
The structure learning framework of RSGNN does not scale to large graphs since it must learn the $n \times n$ entries of the signed adjacency matrix. Indeed,~\citet{zhang2023rsgnn} use a smaller snapshot of Slashdot in their experimental section.

\subsection{Ablation Study}
In this subsection, our aim is to understand how different components of our proposal impact the results. 
Specifically, we ablate three different components: $1)$ the multiscale SB for the \textbf{RQ3}, $2)$ the \texttt{ReWeight} procedure for the \textbf{RQ4}, and $3)$ the amount of unsupervised and supervised data available for the \textbf{RQ5}.

\spara{(RQ3) Microscale and Mesoscale Social Balance are both informative for link polarity prediction.}
Our preliminary hypothesis, as described in the Introduction, is that Multiscale Social Balance is a much stronger concept than considering Micro or Mesoscale Social Balance individually. We now aim to provide empirical validation of this intuition by ablating either the Microscale or the Mesoscale component in the \ourProcedure{} training framework.
\begin{table}[t]
    \centering
    \setlength{\tabcolsep}{3pt} %
    \renewcommand{\arraystretch}{1.1} %
    \fontsize{8}{8}\selectfont %
    \begin{tabular}{lp{15mm}p{12mm}p{15mm}p{15mm}}
         \toprule
         Dataset & Perturb (\%) & L2RW & L2RW-MicroSB & L2RW-MesoSB \\
         \midrule
         \multirow{3}{*}{Bitcoin-Alpha} & 0 & 0.6846 & 0.6814 & $0.6770^*$ \\
         {} & 10 & 0.5897 & $0.5608^*$ & $0.5766^*$ \\
         {} & 20 & 0.5614 & $0.5557^*$ & $0.5340^*$ \\
         \midrule
         \multirow{3}{*}{Bitcoin-OTC} & 0 & 0.7438 & 0.7399 & $0.7368^*$ \\
         {} & 10 & 0.6672 & $0.6478^*$ & $0.6519^*$ \\
         {} & 20 & 0.6337 & $0.6169^*$ & $0.6112^*$ \\
         \midrule
         \multirow{3}{*}{Wiki} & 0 & 0.7634 & $0.7588^*$ & 0.7623 \\
         {} & 10 & 0.7297 & 0.7244 & 0.7240 \\
         {} & 20 & 0.7054 & $0.6973^*$ & 0.7061 \\
         \midrule
         \multirow{3}{*}{Slashdot} & 0 & 0.7520 & $0.7464^*$ & 0.7505 \\
         {} & 10 & 0.6963 & $0.6838^*$ & $0.7004^*$ \\
         {} & 20 & 0.6763 & $0.6635^*$ & 0.6745 \\
         \bottomrule
    \end{tabular}
    \vspace{1mm}
    \caption{Macro-F1 for link polarity prediction by ablating either the Microscale SB (indicated as -MicroSB) or the Mesoscale SB (indicated as -MesoSB). \ourProcedure{} is referred to as "L2RW". The asterisk denotes if the ablation significantly changed the model according to a two-sided paired t-test (* if $<0.05$).}
    \label{tab:ablation-micro-and-meso-sb}
\end{table}
And in fact Tab.~\ref{tab:ablation-micro-and-meso-sb} provides that combining both components is crucial to achieving the best results. Interestingly, there is no clear winner between Microscale and Mesoscale Social Balance, as the former is more important on Wiki and Slashdot, while the latter is more significant on Bitcoin-Alpha and Bitcoin-OTC. This underscores the importance of considering both scales simultaneously to avoid losing critical information. Notably, this empirical evidence seems to confirm the findings in Tabs~\ref{tab:null-model} and ~\ref{tab:null-model-meso}, where the z-scores for microscale SB are higher than those for mesoscale SB in the Wiki and Slashdot datasets, while the opposite is true for Bitcoin-Alpha and Bitcoin-OTC.

\spara{(RQ4) Learning to reweight social balance increments accuracy.}
\begin{table}[t]
    \centering
    \setlength{\tabcolsep}{4pt} %
    \renewcommand{\arraystretch}{1.1} %
    \fontsize{8.3}{8}\selectfont %
    \begin{tabular}{cccc}
         \toprule
         Dataset & Perturb (\%) & \ourProcedure{} & ConstantWeight \\
         \midrule
         \multirow{3}{*}{Bitcoin-Alpha} & 0 & 0.6846 & $0.6732^*$ \\
         {} & 10 & 0.5897 & $0.5728^*$ \\
         {} & 20 & 0.5614 & $0.5376^*$ \\
         \midrule
         \multirow{3}{*}{Bitcoin-OTC} & 0 & 0.7438 & $0.7322^*$ \\
         {} & 10 & 0.6672 & $0.6472^*$ \\
         {} & 20 & 0.6337 & $0.6039^*$ \\
         \midrule
         \multirow{3}{*}{Wiki} & 0 & 0.7634 & 0.7644 \\
         {} & 10 & 0.7297 & 0.7311 \\
         {} & 20 & 0.7054 & 0.7057 \\
         \midrule
         \multirow{3}{*}{Slashdot} & 0 & 0.7520 & $0.7430^*$ \\
         {} & 10 & 0.6963 & $0.6939^*$ \\
         {} & 20 & 0.6763 & $0.6736^*$ \\
         \bottomrule
    \end{tabular}
    \vspace{1mm}
    \caption{Macro-F1 for link polarity prediction by ablating the \texttt{ReWeight} procedure and setting the weight value of each edge in $\mathcal{E}_{\text{SB}}$ equal to 1. The asterisk denotes if the ablation significantly changed the model according to a two-sided paired t-test (* if $<0.05$).}
    \label{tab:ablation-lrw}
\end{table}
\begin{figure*}[ht]
\centering
\begin{tabular}{cccc}
  \includegraphics[width=.49\columnwidth]{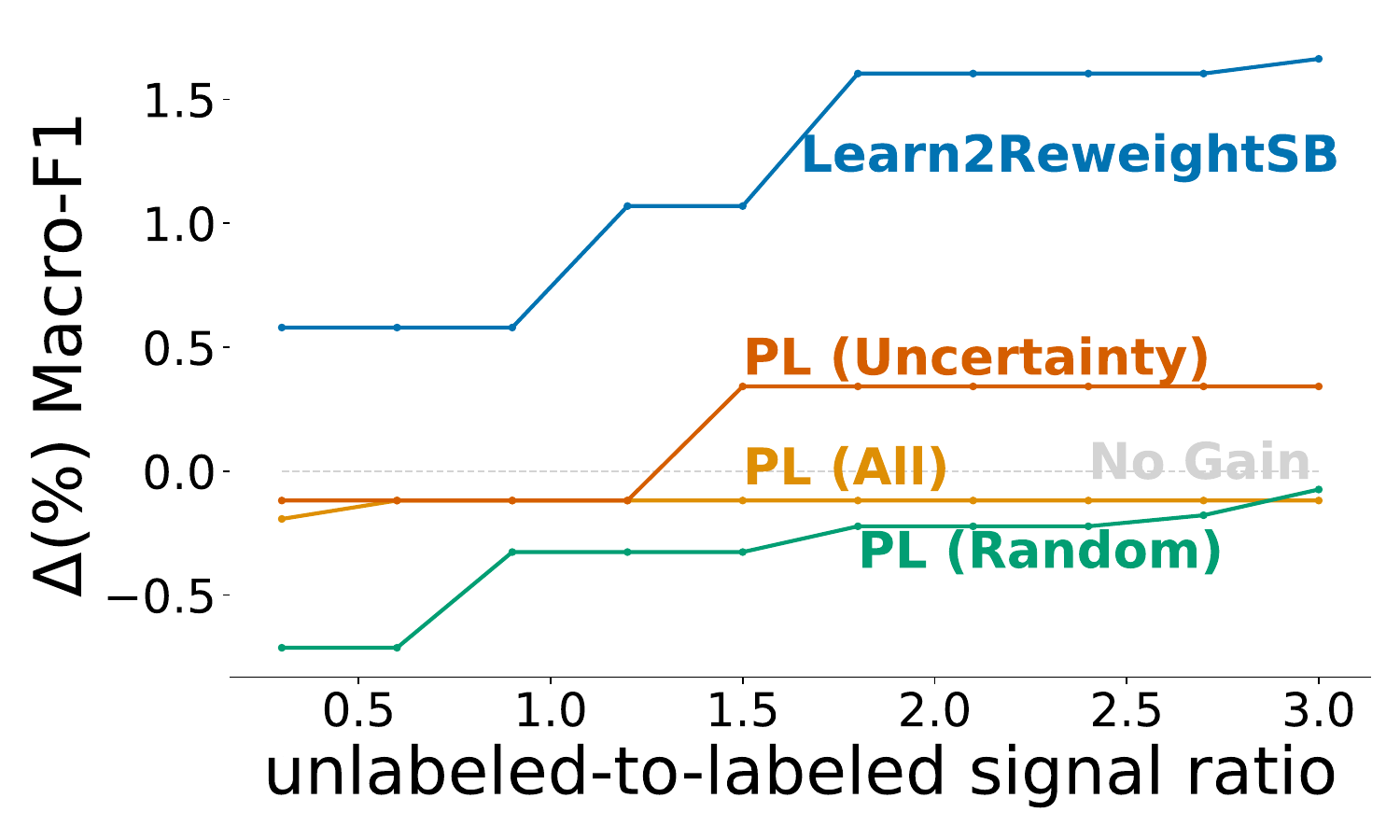} &   \includegraphics[width=.49\columnwidth]{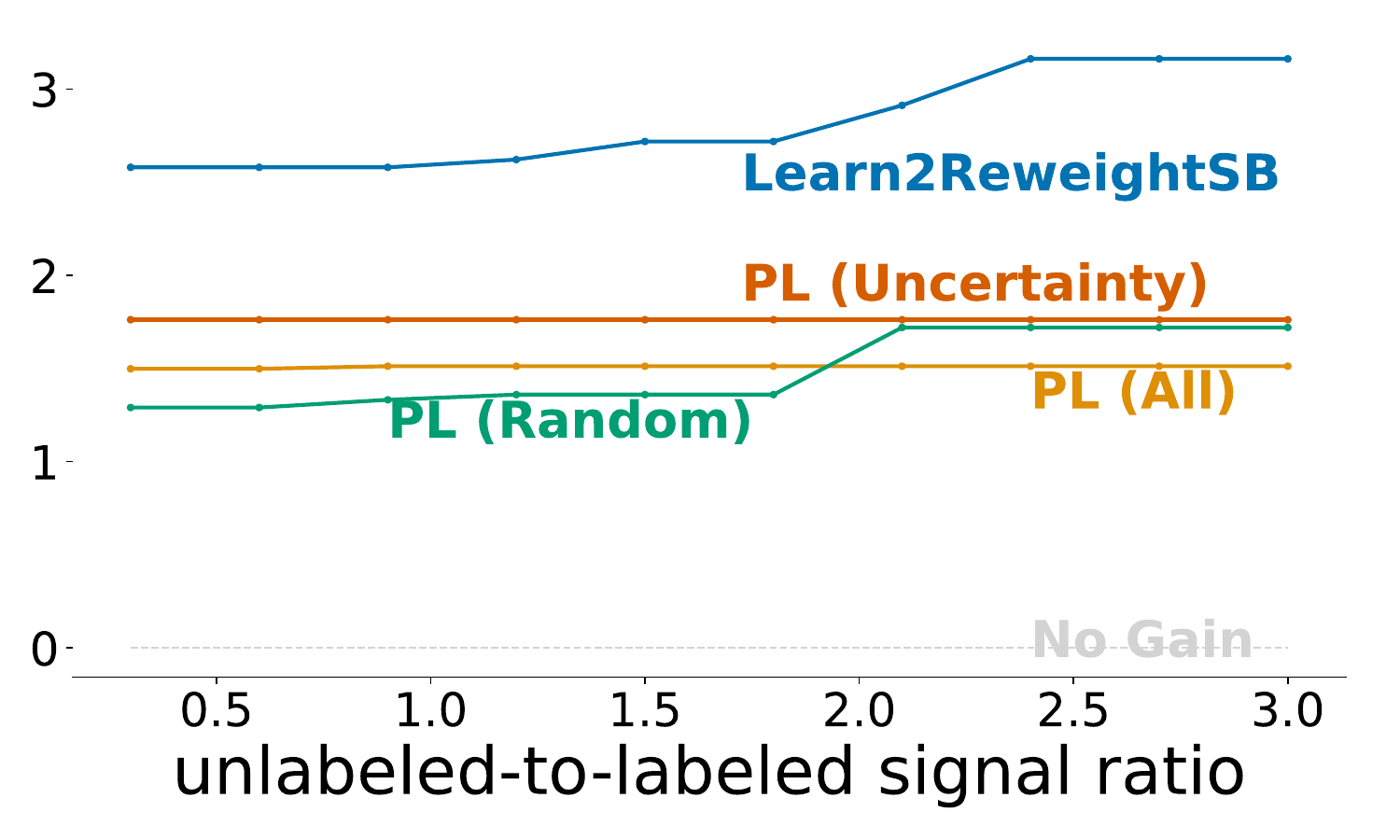} &
  \includegraphics[width=.49\columnwidth]{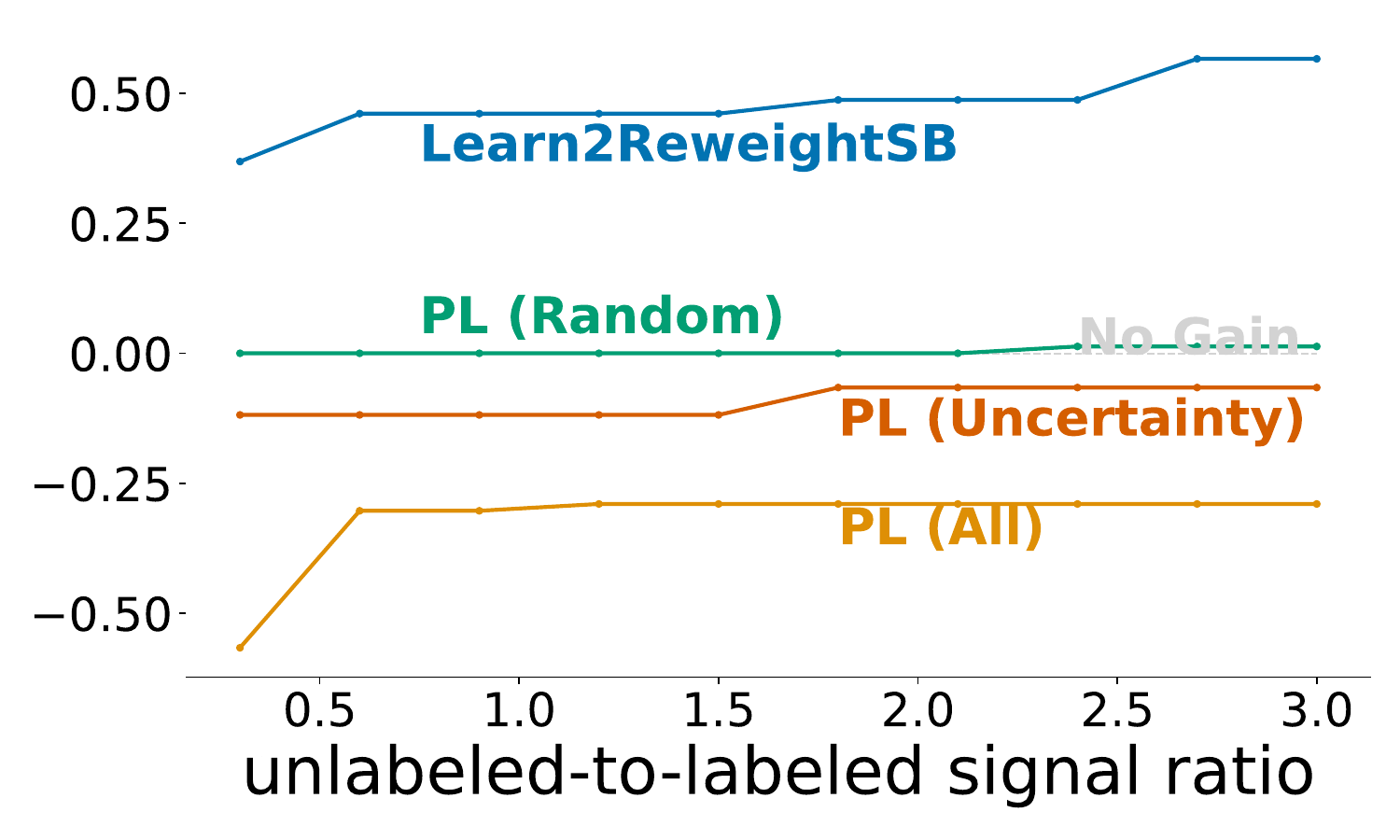} &   \includegraphics[width=.49\columnwidth]{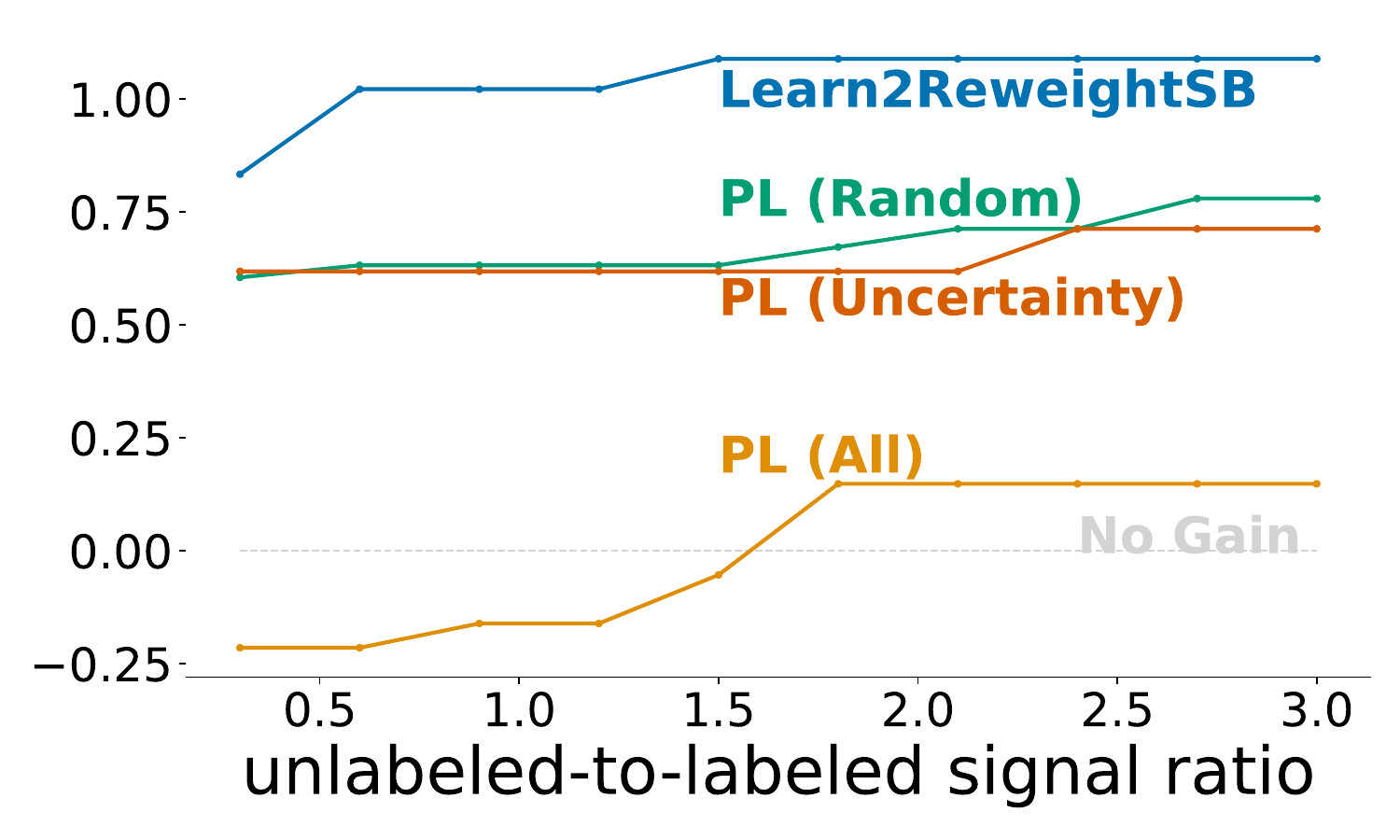} \\
\hspace{0.45cm} (a) Bitcoin-Alpha & \hspace{0.45cm} (b) Bitcoin-OTC  & (c) Wiki & \hspace{0.45cm} (d) Slashdot \\[6pt]
\end{tabular}
\vspace{-5mm}
\caption{
Gain in Macro-F1 for our \ourProcedure{} and competitor methods compared to an SDGNN trained only on $\mathcal{E}^L$, relative to the number of unlabeled edges. The performance of \ourProcedure{} increases gracefully as more unlabeled edges are incorporated into the framework.}
\label{fig:ablation-unsupervised}
\end{figure*}
\begin{figure}[ht]
\begin{tabular}{cc}
  \includegraphics[width=0.45\columnwidth]{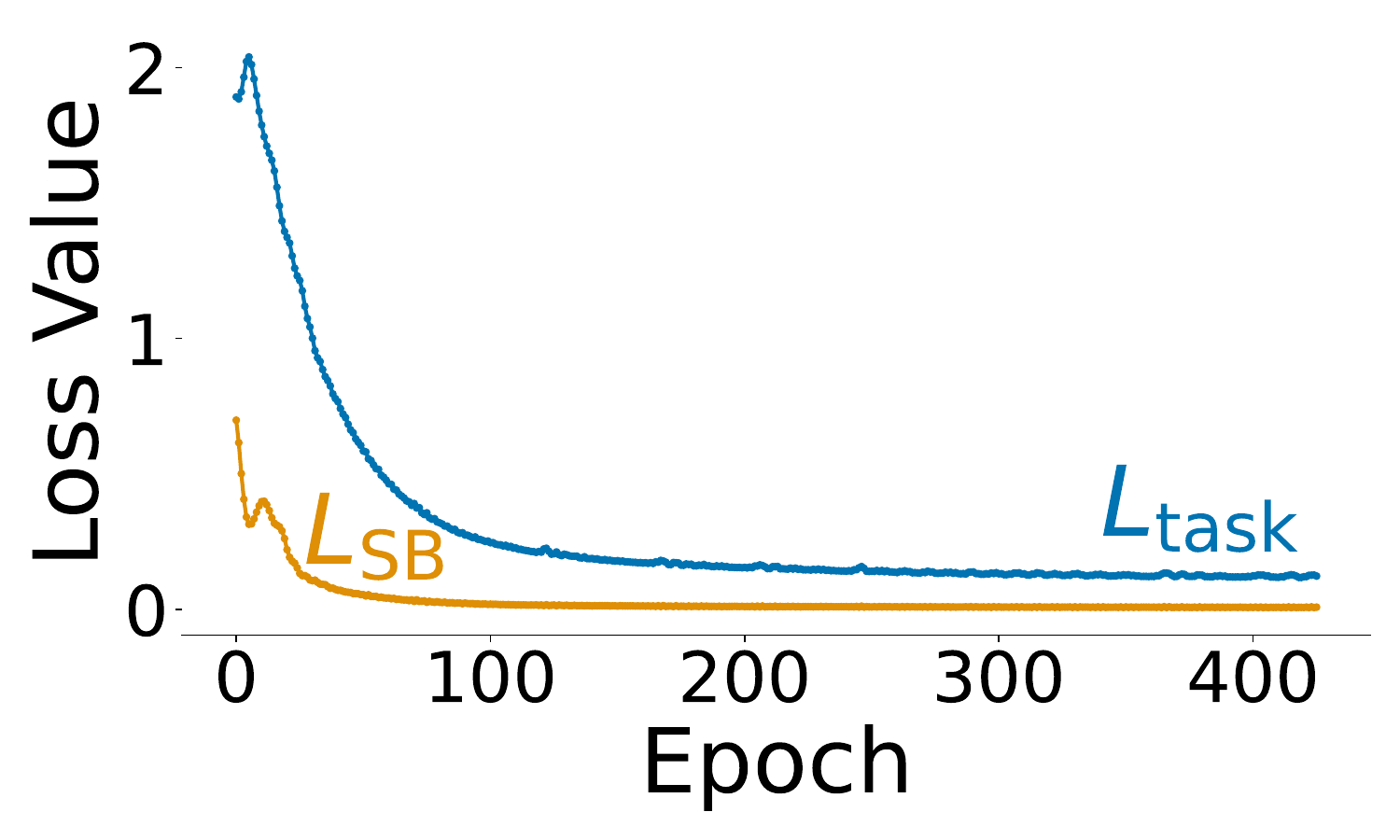} &   \includegraphics[width=0.45\columnwidth]{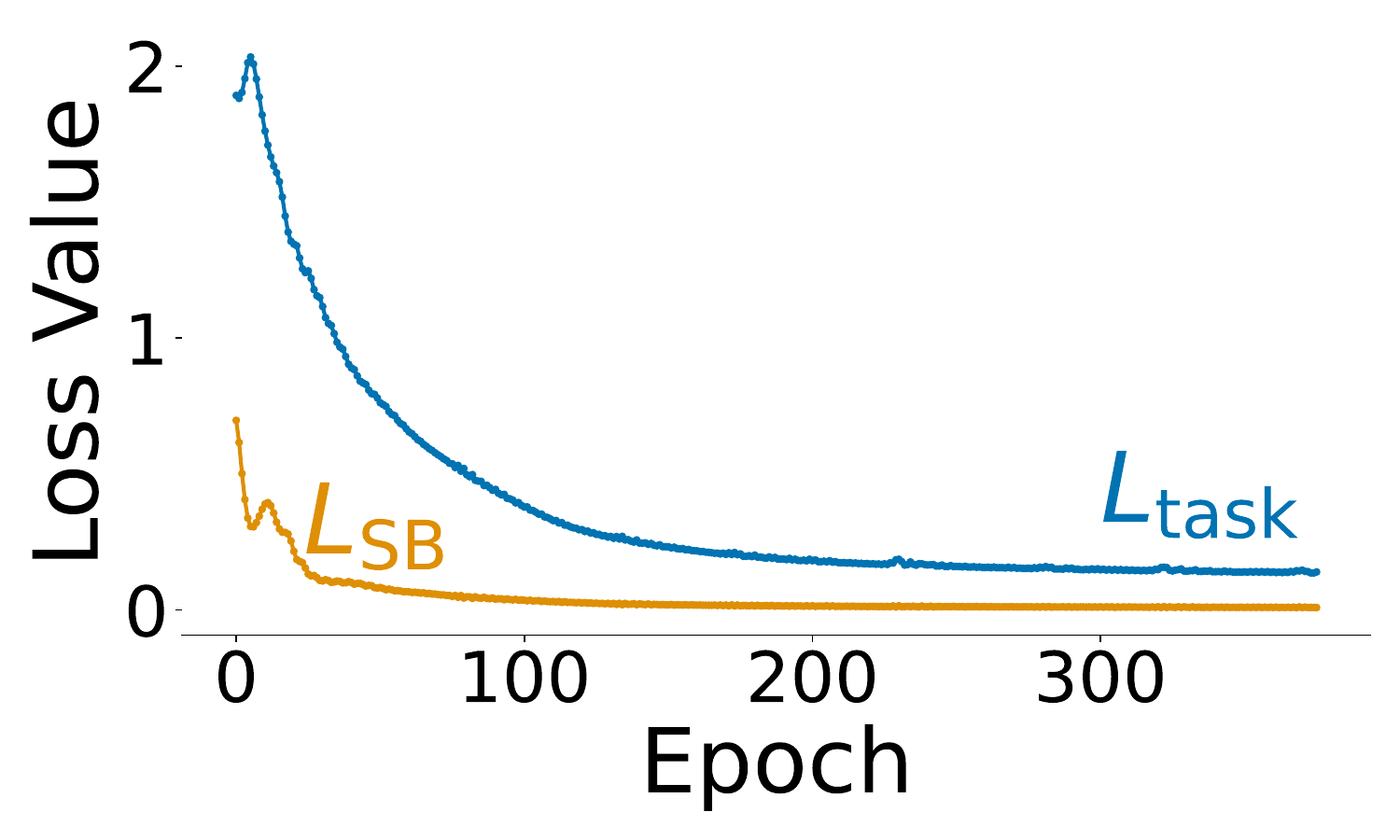} \\
\hspace{0.45cm} (a) Bitcoin-Alpha & \hspace{0.45cm} (b) Bitcoin-OTC \\
\includegraphics[width=0.45\columnwidth]{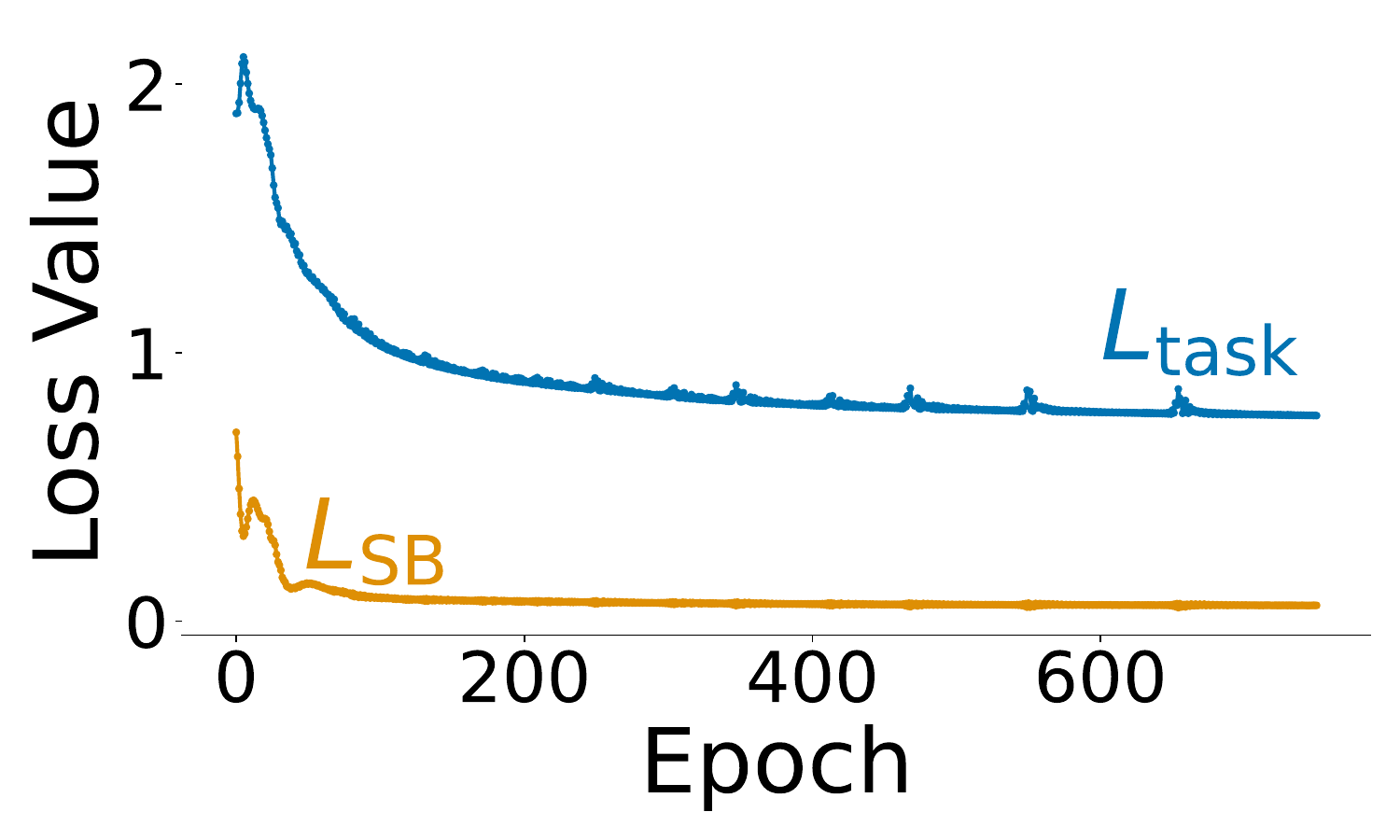} &   \includegraphics[width=0.45\columnwidth]{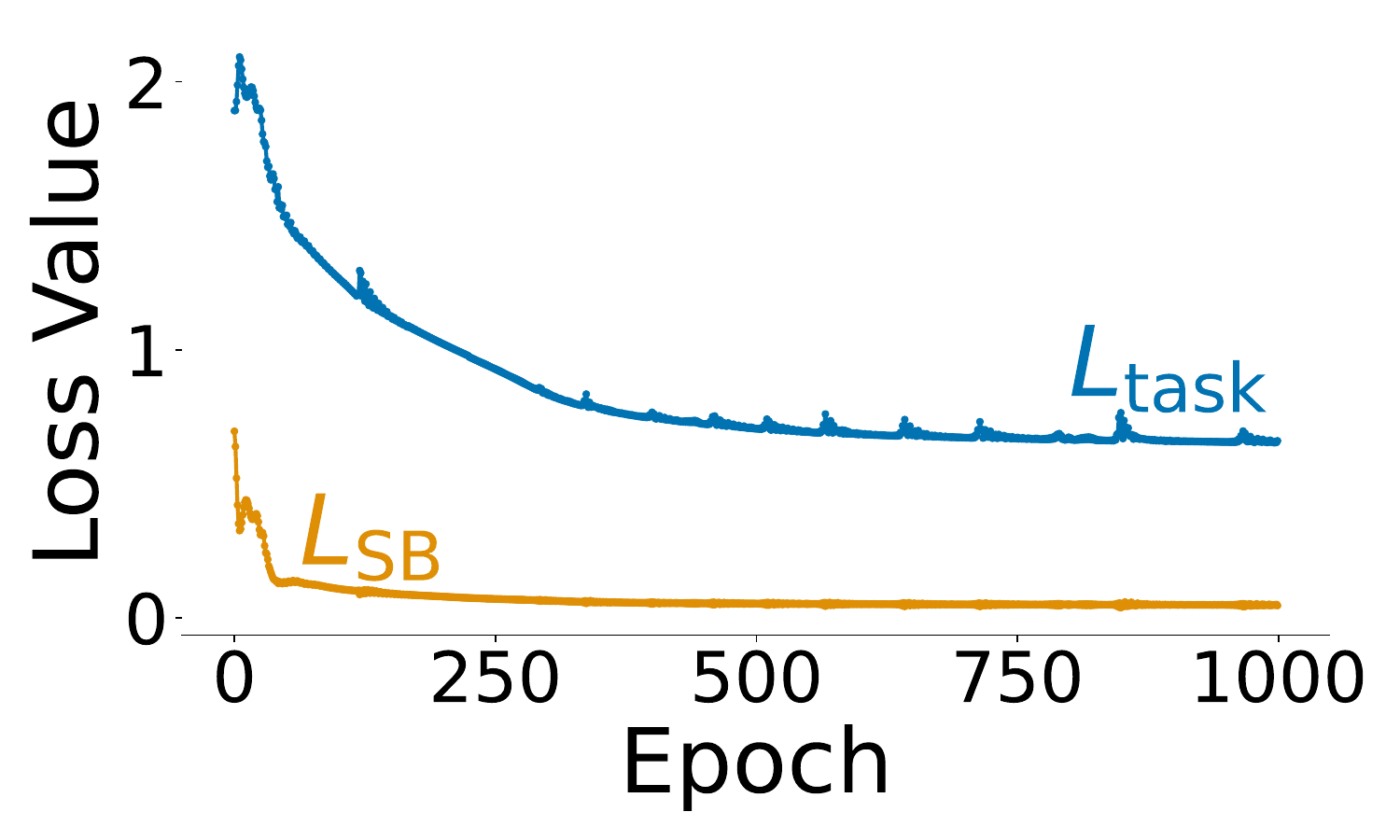} \\
\hspace{0.45cm} (a) Wiki & \hspace{0.45cm} (b) Slashdot \\[6pt]
\end{tabular}
\vspace{-5mm}
\caption{
Learning curves across training epochs of our framework. 
Both $l_{\text{task}}$ and $l_{\text{SB}}$ decrease until reaching a plateau.}
\label{fig:learning-curves}
\end{figure}
Our proposal utilizes Multiscale SB to assign a sign to unlabeled edges, followed by a procedure to adjust for cases where Multiscale SB may fall short. We now ablate this second part of our framework to understand its contribution compared to "blindly" trusting each edge and assigning a weight of 1 to all edges in $\mathcal{E}_{\text{SB}}$.
Tab.~\ref{tab:ablation-lrw} demonstrates that "blindly" trusting all edges labeled with Multiscale SB (i.e., ConstantWeight in the table) is less effective than the \texttt{ReWeight} procedure. Except for Wiki, where the two strategies tie, \ourProcedure{} consistently improves the results, with the performance gap widening as noise is introduced into the data.
Furthermore, we investigate whether \ourProcedure{} converges during the training phase. Fig.~\ref{fig:learning-curves} shows the decrease in both the $l_{\text{task}}$ (blue line) and $l_{\text{SB}}$ (orange line) components across training epochs, eventually reaching a plateau. For brevity, we report the results obtained on the four datasets with no perturbed signs.

\spara{(RQ5) \ourProcedure{} improves with respect to the unlabeled-to-labeled signal ratio.}
In a final set of experiments, we aim to understand how the the amount of injected unlabeled information in the dataset impacts the performance of our proposed framework.
For each dataset, we create nine subsets of the unlabeled edges, $\mathcal{E}_{\text{10\%}}^U \subset \cdots \subset \mathcal{E}_{\text{80\%}}^U \subset \mathcal{E}_{\text{90\%}}^U \subset \mathcal{E}^U$, by down-sampling the unlabeled set $\mathcal{E}^U$ in increments of 10\%. This approach generates nine different views of the datasets with varying degrees of unlabeled information.
The results are presented in Fig.~\ref{fig:ablation-unsupervised}. The x-axis represents the unlabeled-to-labeled signal ratio, i.e., $\frac{\lvert \mathcal{E}_{\text{x\%}}^U \rvert}{\lvert \mathcal{E}^L \rvert}$, for the nine subsets, while the y-axis shows the Macro-F1 performance gain over SDGNN, which uses only supervised information. 
For this analysis, we fix the noise ratio to 0\%.
We observe that the performance of \ourProcedure{} improves as more unlabeled information is added to the dataset. Notably, in the case of the Bitcoin-Alpha dataset, our framework appears to benefit from additional unlabeled edges. By contrast, all other competitors that use unsupervised information seem to reach an early plateau, thus showing no substantial benefit from the inclusion of new unlabeled information.

%% file: conclusions.tex
This study improves the capabilities of SGNNs for link polarity prediction in signed networks, a particularly challenging task for SGNNs when the available data are sparse or noisy.
Our semi-supervised learning framework employs the innovative concept of multiscale social balance, dynamically reweighting data samples based on their assessed correctness and incorporating structural information from unlabeled edges.
Empirical validations confirm that our approach significantly boosts the performance of an input SGNN model.
Moreover, the experiments validate our hypothesis that different datasets can vary significantly in the level of information contained within microscale and mesoscale social balances, making our framework suitable for boosting performance across various settings. 
Finally, we found that learning to reweight the importance of unlabeled data, based on social balance, can overcome the practical limitations of social balance theory when its universal application is challenged.
Future efforts will aim to extend our framework towards several directions of interest. First, it would be interesting to investigate how the framework can adapt to temporal networks, allowing us to capture the dynamic nature of relationships and interactions over time. By incorporating temporal data, we can better model evolving patterns and trends within the network. Another direction for further investigation is to analyse how our framework can be accommodated to also address the inherent sign imbalance that occurs in real data, as shown in the data analysis discussed in Section~\ref{sec:experiments}.
Additionally, we will explore further application tasks such as: link prediction, where the goal is to predict the formation of new links or the dissolution of existing ones, besides the underlying sign;  and community detection, which involves identifying clusters or groups of nodes with dense sign-consistent interconnections. 
These extensions will not only broaden the applicability of our framework but also enhance its robustness and utility in real-world scenarios.